\input{aipcheck}

\documentclass[
    ,final            
  ]
  {aipproc}

\usepackage{pstricks,pst-node,pst-text,pst-3d}
\usepackage{amsmath}
\usepackage{amssymb}
\usepackage{curves}
\usepackage{shadow}
\usepackage{feynmf}

\layoutstyle{8x11single}

\def\Journal#1#2#3#4{{#1} {\bf #2}, #3 (#4)}

\def\etal{{\it et al.}}

\def\APJ{\em ApJ.}
\def\APP{\em Astropart. Phys.}

\def\JCA{\em J. Cosmol. Astrop. Phys.}

\def\NPB{{\em Nucl. Phys.} B}

\def\PLB{{\em Phys. Lett.}  B}

\def\PRD{{\em Phys. Rev.} D}
\def\PRL{\em Phys. Rev. Lett.}

\def\RPP{\em Rept. Prog. Phys.}

\def\be{\begin{equation}}
\def\ee{\end{equation}}
\def\bea{\begin{eqnarray}}
\def\eea{\end{eqnarray}}
\def\bes{\begin{equation*}}
\def\ees{\end{equation*}}
\def\beas{\begin{eqnarray*}}
\def\eeas{\end{eqnarray*}}

\def\mg{\mathsf g}
\def\cx{\mathtt X}
\def\ca{\mathtt A}

\def\phix{{\phi}_x}

\begin{document}

\title {Dark energy and formation of classical scalar fields}


\author{Houri Ziaeepour}
{
address={Mullard Space Science Laboratory, University College London\\
Holmbury St. Mary, Dorking, Surrey, RH5 6NT, UK.\\ Email: {\tt hz@mssl.ucl.ac.uk}}
}

\begin{abstract}
We present a quintessence model for the dark energy in which the quintessence 
scalar field is produced by the decay of a super heavy dark matter and 
gradually condensate to a classical scalar field. This model can explain both 
the smallness and the latest observations by WMAP for the equation of state 
of the dark energy which has $w \sim -1.06$. We review both classical and 
field theoretical treatment of this model and briefly explain the most 
important parameters for obtaining the observed characteristic of the dark 
energy.
\end{abstract}

\maketitle
\begin{fmffile}{fmfgenfydig}
\section{Introduction} \label{sec:intro}
Quintessence models are one the most popular candidates for the dark energy. 
The main ingredient of these models is a classical scalar field. It is not 
difficult to find scalar particles/quantum fields in the Standard Model and 
its extension, and in cosmological context. But it is evident to find a field 
with necessary characteristics. Some authors have suggested the same field as 
the driver force behind inflation and dark energy\cite{infquin}. The common 
aspect of these models is a non-zero
slowly decreasing classical potential during inflation era which at late time 
settles to a very small value and behaves as the dark energy today. Even 
without direct relation to inflaton, most of other quintessence models also 
predict a late time tracker classical scalar field with a potential which in 
one way or another gets a small but non-zero value at late times. As for the 
particle physics of these models, all extensions of the Standard 
Model include various types of scalar fields, from very light ones such as 
QCD axion, dilaton, and modulies in string theories, to presumably heavy 
particles such as super partner of spinor in supersymmetry models and Higgs 
bosons. 

There are however one problematic issue regarding the mass of quintessence 
fields. In most quintessence models as the classical scalar field is related 
to the phenomena at very high energies, it is difficult 
to make its value very small without extreme fine-tuning. For this reason its 
mass and coupling must be small, in most cases of order $\sim 10^{-32} eV$ or 
it must have tachyonic potential - a potential with a negative effective mass 
such that the mass and interaction term cancel each other. 
It is not easy to incorporate such a small mass into particle physics models. 
In most cases radiative corrections make the bare small mass much larger than 
what is needed\cite{massrenor}. Exceptional cases exist in which due to a 
translation symmetry a pseudo-Nambu-Goldstone boson does not get correction 
after renormalisation and can keep is smallness\cite{quinpngb}. The form of 
their potentials is not however very simple and it is not sure 
if they can be easily obtained from particle physics model without fine-tuning 
or additional concepts which are yet more sophisticated. This type of 
modelling just shift an unexplained problem to another one and does not seem 
to be a concrete solution of the problem. In addition, the latest results 
from WMAP\cite{wmap} shows that the simplest inflation 
model with just a mass term in the potential is the best fit to the CMB 
anisotropy data. If this conclusion is confirmed, models which try to unify 
quintessence field with inflaton would be compromised or at least will have 
more difficulties to explain the late behaviour of this field. It would not 
be easy to explain a very small but non-zero value for the field with such a 
simple potential unless other ingredients and most probably some fine-tuning 
be added to the models.

There is also an unexplored issue regarding the formation of a classical 
scalar field and its potential. Although a large number of quantum scalar 
fields are predicted by particle physics models, their condensation to a 
classical field is not a trivial process specially when complex potentials are 
requested for quintessence models. The importance of condensation is not  
limited to the quintessence models of dark energy and is an indispensable 
content of Higgs phenomena, symmetry breaking and phase transition in the 
context of baryo and lepto genesis, electroweak and QCD. 

Here we first review a quintessence model based on the 
condensation of a light scalar produced by slow decay of a heavy dark matter 
and show that it can explain the observation of $w \lesssim -1$ for the 
equation of state of the dark energy. We also review the methodology and some 
analytical results of the field theoretical study of the condensation 
process and discuss important parameters for its formation and evolution.

\section{A meta-stable super heavy dark matter}
The motivation for this type of dark matter is the observation of Ultra High 
Energy Cosmic Rays (UHECR)\cite{uhecr}. Acceleration models have great 
difficulties to produce enough particles at these huge energies - larger 
than $\sim 10^{20}eV$. 
Moreover, the sites of the potential accelerators, magnetars and AGNs, which 
are marginally capable of accelerating charged particles to such energies 
must be close - in a distance $< 50~Mpc$ - otherwise the interaction with the 
CMB decelerates nucleons. This process should make what is called GZK cutoff 
in the high energy tail of the cosmic rays spectrum and has been predicted in 
1960s, but is not observed. Considering the small fluxes which can be 
provided by each source at these energies, it does not seem that there are 
enough sources in the limited volume space permitted by the CMB above to 
explain the observed spectrum by standard Fermi acceleration or its 
improvement by taking into account non-linear effects e.g Alfven shocks. 

Top-down models\cite{topdown} as the origin of UHECR considers the decay of a 
meta-stable 
super heavy particle particle (here on called $X$) as the sources of 
(anti)-nucleons producing the high energy shower in the Earth atmosphere. In 
this case the contribution of the halo of the Galaxy is much more important 
than cosmological ones\cite{xx}\cite{houriwimpzilla}. A detail calculation of 
energy dissipation of high energy 
particles show that for $m_x \sim 10^{22} - 10^{24} eV$ and lifetimes even as 
short as $\sim 10\tau_0$ where $\tau_0$ is the present age of the Universe, 
roughly the totality of dark matter can be $X$ particles without violating 
the present observations of the UHECR flux\cite{houriwimpzilla}.

There are a number of particle physics candidates for $X$ particles. In models 
inspired by string theory and various possible compactifications, meta-stable 
particle called {\it crypton} with a mass $\sim 10^{22} eV$ or higher can be 
$X$ particles\cite{crypton}. Observation of neutrino oscillation and the 
limit on their masses $\sum m_{\nu} < 0.68 eV$\cite{wmap} along with the close 
to  degeneracy of their mass difference $\Delta m^2 \approx 10^-3$ makes 
seesaw mechanism the most favourite model for explaining these observations. 
In this case, the small mass of left neutrinos push the mass of 
right hand neutrinos to $~10^{23} eV$ or higher. In supersymmetric 
models seesaw can also exist in the super-partner sector of neutrinos and if 
some of right hand sneutrinos are meta-stable due to discrete global 
symmetries, they can be also good candidates for $X$ particles\cite{rnu}. 

Besides explaining the mystery of UHECRs, a meta-stable dark matter can 
also solve another mystery: a $w \lesssim -1$ for the dark energy where
$w = p/\rho$ with $p$ and $\rho$ respectively pressure and density of dark 
energy. An ordinary quintessence model based on a scalar field with positive 
kinetic energy has always $w \geqslant -1$ and therfore can not explain the 
observed equation of state of the dark energy if $w \lesssim -1$ is 
confirmed. Nonetheless, it is 
possible to show that in presence of cosmological constant i.e. when $w = -1$ 
and a slowly decaying\cite{houriustate} (or interacting\cite{quindmint}) dark 
matter, if supernovae data is analysed with prior assumption of stable dark 
matter, an effective $w \leqslant -1$ will be obtained. To prove this, we 
use an approximation solution of the expansion equation:
\be
H^2 (z) = \frac {8\pi G}{3} T^{00} (z) + \frac {\Lambda}{3} \label {expeq}
\ee
where H is the Hubble constant and $T^{\mu\nu}$ is the energy-momentum tensor 
of the Universe and $T^{00} = \rho$. We want to find an analytical expression 
for the equivalent quintessence model with a stable dark matter to a 
cosmology with a decaying dark matter of lifetime $\tau$ and a cosmological 
constant $\Lambda$. With a good precision the total density of such models 
can be written as the following:
\be
\frac {\rho (z)}{{\rho}_c} \approx {\Omega}_M (1 + z)^3 \exp 
(\frac {{\tau}_0 - t}{\tau}) + {\Omega}_{Hot} (1 + z)^4 + 
{\Omega}_M (1 + z)^4 \biggl (1 - \exp (\frac {{\tau}_0 - t}{\tau}) \biggr ) + 
{\Omega}_{\Lambda}. \label {totdens}
\ee
We assume a flat cosmology i.e. ${\Omega}_M + {\Omega}_{\Lambda} = 1$ 
(ignoring the hot part). ${\rho}_c$ is the present critical density. If the 
dark matter is stable and we neglect the contribution of hot component, the 
expansion factor $a (t)$ is:
\be
\frac {a (t)}{a ({\tau}_0)} = \biggl [ \frac {(B \exp (\alpha (t - 
{\tau}_0)) - 1)^2}{4AB \exp (\alpha (t - {\tau}_0))}\biggr ]^{\frac {1}{3}} 
\equiv \frac {1}{1 + z}. \label {at}
\ee
\be
A \equiv \frac {{\Omega}_{\Lambda}}{1 - {\Omega}_{\Lambda}} \quad , \quad
B \equiv \frac {1 + \sqrt {{\Omega}_{\Lambda}}}{1 - 
\sqrt {{\Omega}_{\Lambda}}} \quad , \quad \alpha \equiv 3 H_0 
\sqrt {{\Omega}_{\Lambda}}
\ee
Using \eqref {at} as an approximation for $\frac {a (t)}{a ({\tau}_0)}$ when 
dark matter decays, \eqref {totdens} takes the following form:
\bea
\frac {\rho (z)}{{\rho}_c} & \approx & {\Omega}_M (1 + z)^3 C^{-\frac {1}
{\alpha \tau}} + {\Omega}_{Hot} (1 + z)^4 + {\Omega}_M (1 + z)^4 (1 - 
C^{-\frac {1}{\alpha \tau}}) + {\Omega}_{\Lambda} \label {totdens1}\\
C & \equiv & \frac {1}{B} \biggl (1 + \frac {4A}{(1 + z)^3} - \sqrt {(1 + 
\frac {4A}{(1 + z)^3})^2 - 1} \biggr)
\eea
where $\alpha \tau \gg 1$ and \eqref {totdens1} becomes:
\bea
\frac {\rho (z)}{{\rho}_c} & \approx & {\Omega}_M (1 + z)^3 + {\Omega}_{Hot} 
(1 + z)^4 + {\Omega}_q (1 + z)^{3 {\gamma}_q}, \label {totdens2} \\
{\Omega}_q (1 + z)^{3 {\gamma}_q} & \equiv & {\Omega}_{\Lambda} (1 + 
\frac {{\Omega}_M}{\alpha \tau {\Omega}_{\Lambda}} z (1 + z)^3 \ln C)
\label {qeqdef}
\eea
Equation \eqref {qeqdef} is the definition of equivalent quintessence model. 
After linearising this expression we find:
\be
w_q \equiv {\gamma}_q - 1 \approx \frac {{\Omega}_M (1 + 4 A)(1 - \sqrt {2 A})}
{3 \alpha \tau {\Omega}_{\Lambda} B} - 1 \label{equivq}
\ee
Using \eqref {equivq}, one can show that if $w_q < -1$ if 
${\Omega}_{\Lambda} > \frac {1}{3}$. In the general case of an interaction 
between dark matter and dark energy, the exponential term in \eqref{totdens} 
can be replaced by a more general expression $f (z)$ which presents the 
modification of the density of both component due to their interaction. It has 
been shown\cite{quindmint} that this function can be selected in such a way 
that $w_q < -1$.

A more precise treatment of this model needs the simulation of the decay and 
the dynamics of of the remnant. This has been done for various values of the 
cosmological parameters as well as mass and lifetime of the dark 
matter\cite{houriustate} and the results has been used to fit this model 
to supernovae data. The best fit of this model for fixed $\tau = 5 \tau_0$ 
predicts following values for other parameters: 
\be
H_0 = 68.4 \quad , \quad \Omega_{\Lambda} = 0.73, 
\quad \Omega_q = 0.69 \quad , \quad w = -1.066  \label {wobs}
\ee
These values are very close to the latest results from combined 
WMAP, SNLS and SDSS. The importance of this result is 
specially in the fact that the value for $\tau$, the lifetime of the dark 
matter, comes from a completely different data set i.e. the flux of UHECR. 
This parameter is very important for the effect of the decay on the equation 
of state of the dark energy. For instance, if $\tau = 50\tau_0$, $w = -1.006$ 
i.e. much closer to the cosmological constant\cite{houriustate}. 
Therefore, this model has not been fine-tuned to obtain the observed value of 
$w$ as most of other models for having $w \lesssim -1$ (what is called 
phantom models) are.

\section{Dark energy from a decaying dark matter} \label{sec:dmdecay}
The question that arises now is whether dark energy can be related to the 
decay of the dark matter too. The reason for such idea is {\it the 
coincidence problem}. We don't have any explanation why the energy 
density of the dark matter and dark energy is so highly fine-tuned in the 
early universe - by a factor of $\sim 10^{30}$ or more, depending on the 
time of their production - such that they become comparable only after 
galaxy formation. The most trivial answer to this question is evidently that 
somehow there is a relation between these two main contents of the 
Universe\cite{interac0}. The nature of their relation however should be in 
such a way that the mentioned {\it fine-tuning} appears as natural. 

In the models with meta-stable, if the dark energy or more 
precisely the quintessence field is produced by the slow decay of the dark 
matter and if it is produced in enough small amount, its density can be small 
and its relation with the dark matter solves the coincidence 
problem\cite{houridmquin}. We 
should also remark that when we say {\it it must be produced in small amount}, 
this looks like a fine-tuning which we want to avoid. However, here the 
branching ratio of the heavy dark matter to dark energy is orders of magnitude 
larger than in usual quintessence models. The reason is the long life of 
the dark matter. Observationally speaking, the estimation of these quantities 
come from completely different data sets - the lifetime of the dark matter 
and the range of its mass is set by the observed flux of the ultra high 
energy cosmic rays. As we discussed in the previous section, in the top-down 
models the contribution of the Galactic halo is much larger than cosmic one. 
It is a local 
process and therefore independent of the cosmological parameters. The 
smallness of dark energy is thus partially due to the long lifetime of the 
dark matter. Below we will see that branching ratio of dark matter to dark 
energy becomes similar to other weak interactions and can be studies in the 
frame of present particle physics models such as supersymmetry and 
supergravity.

In a simple realization of such a model the Lagrangian has the following form:
\be
{\mathcal L} = \int d^4 x \sqrt{-g} \biggl [\frac {1}{2} g^{\mu\nu} 
{\partial}_{\mu} \phix {\partial}_{\nu} \phix + \frac {1}{2} g^{\mu\nu} 
{\partial}_{\mu} \Phi {\partial}_{\nu} \Phi - V (\phix, \Phi, A) 
\biggr ] + {\mathcal L}_A \label {lagrange}
\ee
$\phix$ presents the meta-stable dark matter and $\Phi$ 
the quintessence. For simplicity we assume that it is a scalar, but general 
aspects of the model will be the same if it is a spinor. The field $A$ 
presents collectively other fields. The term $V (\phix, \Phi, A)$ 
includes all interactions including self-interaction potential for 
$\phix$ and $\Phi$:
\be
V (\phix, \Phi, J) = V_q (\Phi) + V_x (\phix) + g {\phix}^m {\Phi}^n + 
W (\phix, \Phi, J) \label {potv}
\ee
The term $g {\phix}^m {\Phi}^n$ is responsible for the annihilation of $X$ 
and back reaction of quintessence field on the dark matter. The potential 
$W (\phix, \Phi, A)$ presents interactions which contribute to the 
decay of $X$ to light fields and to $\Phi$ (in addition to what is shown 
explicitly in \eqref{potv}). The very long lifetime of $X$ 
constrains this term and $g$. They must be strongly suppressed. 
For $n = 2$ and $m = 2$, the interaction term contributes to the 
mass of $\phix$ and $\Phi$. Because of the huge mass of $\phix$ (which 
must come from another coupling) and its very small occupation number:
\be 
<{\phix}^2> \sim 2 {\rho}_x / {m_x}^2 \label {classapprox}
\ee
For sufficiently small $g$ the effect of this term on its mass is very small. 

We assume that $X$ particles don't have self-interaction i.e. 
$V_x (\phix) = 0$ and $V_q (\Phi)$ is a polynomial, in simplest case 
containing only a mass term and $\Phi^4$ self-interaction with a coupling 
$\lambda$. At classical level we replace $\phix$ and $A$ with their density 
using \eqref{classapprox}. With these simplifications, the evolution 
equations for the classical components are:
\bea
\dot{\varphi} [\ddot{\varphi} + 3H \dot{\varphi} + {m_{\Phi}}^2 \varphi + 
\lambda {\varphi}^3] & = & -2g \dot{\varphi}\varphi 
\biggl (\frac {2 {\rho}_x}{{m_x}^2}\biggr ) + {\Gamma}_{\Phi}{\rho}_x 
\label {phiqe} \\
\dot {{\rho}_x} + 3H {\rho}_x & = & - ({\Gamma}_{\Phi} + {\Gamma}_A)
{{\rho}_x} - {\pi}^4 g^2 \biggl (\frac {{{\rho}_x}^2}{{m_x}^3} - 
\frac {{{\rho}_q}'^2}{{m_q}^3}\biggr ) \label {xeq} \\
\dot {{\rho}_A} + 3H ({\rho}_A + P_A) & = & {\Gamma}_A {{\rho}_x} 
\label {jeq} 
\eea
where $\varphi$ is the classical component of $\Phi$ and ${\Gamma}_{\Phi}$ and 
${\Gamma}_A$ are the decay width of $X$ particles to $\Phi$ and $A$ 
respectively.

\begin{figure}
\includegraphics[height=.3\textheight,angle=-90]{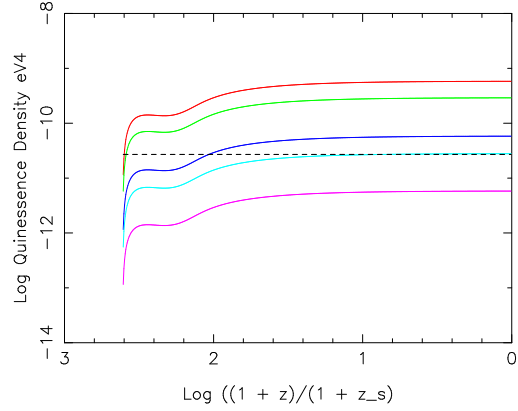}
\caption{Evolution of quintessence field density for ${\Gamma}_0 = 1/\tau_0 
\equiv {\Gamma}_{\Phi}/\Gamma = 10^{-16}$ (magenta), 
$5 {\Gamma}_0$ (cyan), $10 {\Gamma}_0$ (blue), $50 {\Gamma}_0$ (green), 
$100 {\Gamma}_0$ (red). Dash line is the observed value of the dark energy. 
$m_{\Phi} = 10^{-6} eV$, $\lambda = 10^{-20}$. \label{fig:qdens}}
\end{figure}

\begin{figure}
(a)\includegraphics[height=.3\textheight,angle=-90]{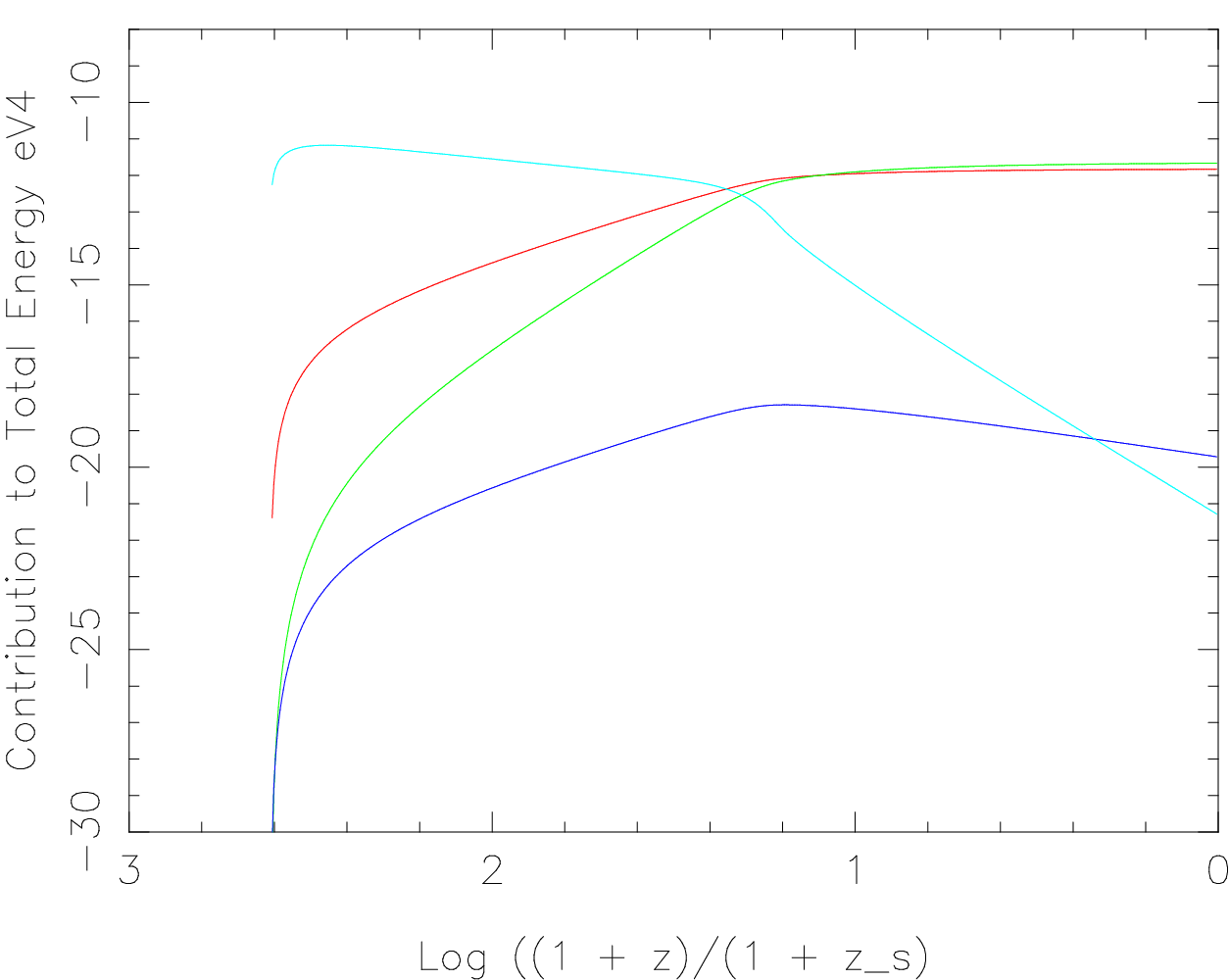}
(b)\includegraphics[height=.3\textheight,angle=-90]{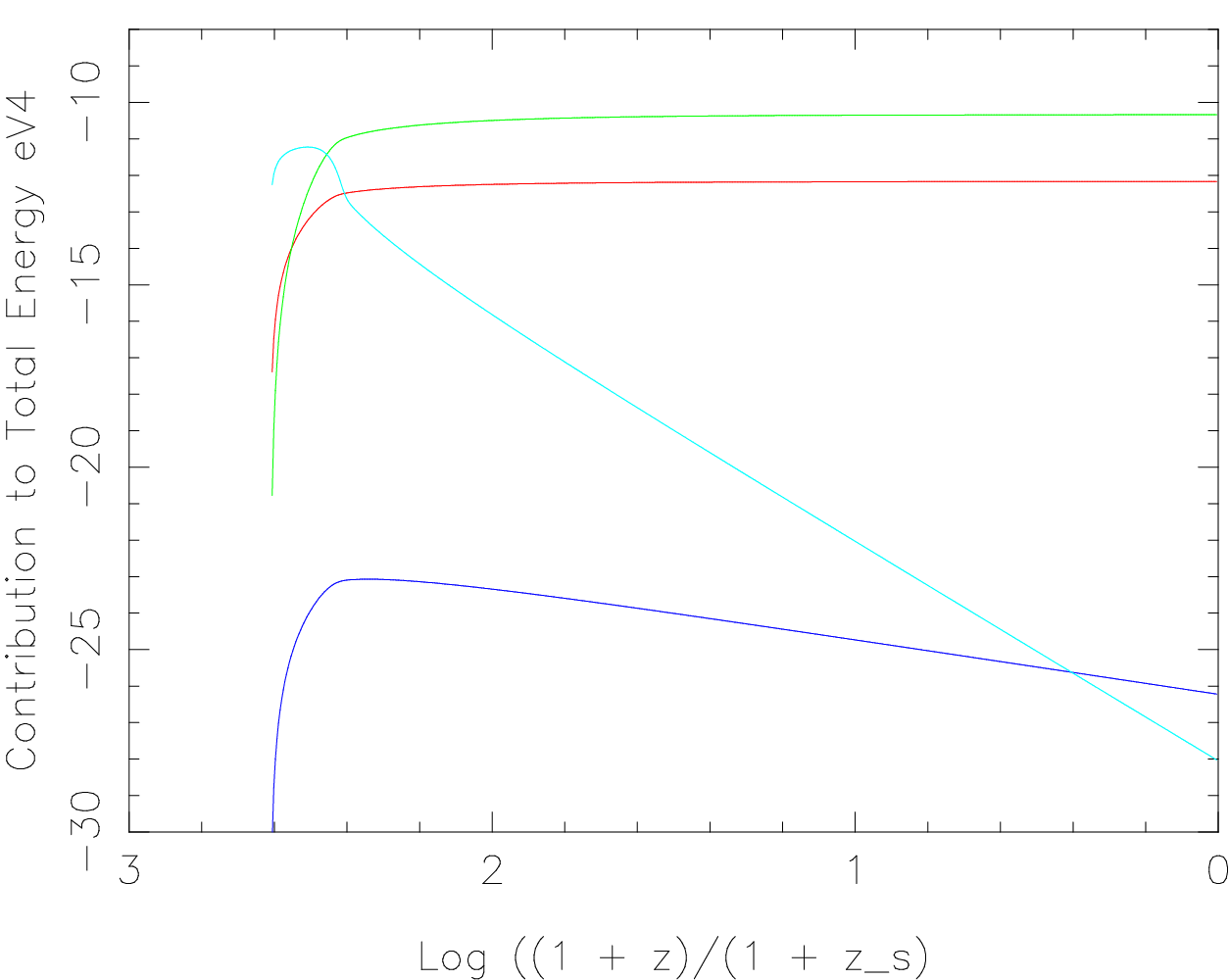}
\caption{Evolution of the contribution to the energy density of 
$\varphi$ for ${\Gamma}_0 \equiv {\Gamma}_q/\Gamma = 10^{-16}$ and : Left, 
$m_q = 10^{-8} eV$ and $\lambda = 10^{-20}$; Right, $m_q = 10^{-6} eV$ and 
$\lambda = 10^{-10}$.Curves are: mass (red), self-interaction (green), 
kinetic energy (cyan) and interaction with super heavy dark matter (blue). 
\label{fig:diffparam}}
\end{figure}

Numerical solution of these equations along with Einstein equations for 
typical values of the parameters show that in such models the classical field 
behaves like a cosmological constant - after few orders of magnitude in 
redshift from the end of $X$ production, presumably after reheating, its 
density approaches a constant and it keeps this behaviour until present, see 
Fig.\ref{fig:qdens}. Variation of parameters by few orders of magnitudes 
does not change the general aspect of $\varphi$ evolution, see 
Fig.\ref{fig:diffparam}-(a) and (b). Therefore this model is not fine-tuned. 

\section {Formation of a classical field} \label{sec:classfield}
The results of the previous section are obtained for a classical field. In the 
Nature however all the particles and fields have quantum behaviour. 
Decoherence 
can make particles to behave semi-classically, i.e. have a wave function 
restricted spatially and in momentum space. A classical field however is more 
complex and classical behaviour of particles does not necessarily means a 
collective behaviour as a classical field. Therefore, in a more rigorous 
treatment of the model of the previous section, we must begin with the 
corresponding quantum field theory, calculate quantum expectations and see 
if these expectations which are equivalent to the classical fields behave as 
requested for a dark energy.

The classical field is defined as the expectation value of the quantum field:
\be
\varphi (x) \equiv \langle \Psi|\Phi (x)|\Psi\rangle \label{classphi}
\ee
where $|\Psi\rangle$ is an element of the Fock space of the system. It is also 
called a {\it condensate} in analogy with quantum mechanics. Using 
canonical representation of the field $\Phi$, it is easy to see that for a 
free quantum scalar field $\langle \Psi|\Phi|\Psi \rangle = 0$. Therefore, 
a necessary condition for appearance of a classical scalar field is 
interaction. The detail study of the Lagrangian \eqref{lagrange} needs an 
explicit definition of interaction/decay term. We consider following decay 
modes for $X$ particles:

\vspace{1cm}
\bea
(a) \quad \quad \quad \quad \quad \quad \quad & \quad \quad \quad 
\quad \hspace {1cm}\text {or}~~ (b) \nonumber \\
\begin{fmfgraph*}(30,30)
\fmfleft{i1}
\fmfright{o2,o3,o4}
\fmf{dbl_plain_arrow,label=$X$}{i1,v1}
\fmf{fermion,label=$A$}{v1,o2}
\fmf{dashes}{v1,v2}
\fmf{plain_arrow,label=$\Phi$}{v2,o3}
\fmf{fermion,label=$A$}{v1,o4}
\fmfdot{v1}
\end{fmfgraph*} & \quad \quad&  \hspace {1cm} 
\begin{fmfgraph*}(30,30)
\fmfleft{i1}
\fmfright{o2,o3,o4}
\fmf{dbl_plain_arrow,label=$X$}{i1,v1}
\fmf{plain_arrow,label=$\Phi$}{v1,o2}
\fmf{plain}{v1,v2}
\fmf{fermion,label=$A$}{v2,o3}
\fmf{plain_arrow,label=$\Phi$}{v1,o4}
\fmfdot{v1}
\end{fmfgraph*} \label{decaymode}
\eea
\end{fmffile}
By considering two modes we can investigate how the type of interactions 
in the model can change the outcome for the classical field.

After adding the corresponding interaction term to the Lagrangian and 
neglecting the quadratic interaction between $X$ and $\Phi$, we find 
following equations for the evolution of the classical field $\varphi$:
\bea
\frac{1}{\sqrt{-g}}{\partial}_{\mu}(\sqrt{-g} g^{\mu\nu}{\partial}_{\nu}
\varphi) + m_{\Phi}^2 \varphi + \frac{\lambda}{n}\sum_{i=0}^{n-1} (i+1)
\binom{n}{i+1}{\varphi}^i\langle{\phi}^{n-i-1}\rangle - \mg \langle 
XA^2\rangle = 0 && \nonumber \\
\hspace {12cm} \text{For (\eqref{decaymode})-a} && \label {dyneffa} \\ 
\frac{1}{\sqrt{-g}}{\partial}_{\mu}(\sqrt{-g} g^{\mu\nu}{\partial}_{\nu}
\varphi) + m_{\Phi}^2 \varphi + \frac{\lambda}{n}\sum_{i=0}^{n-1} 
(i+1)\binom{n}{i+1}{\varphi}^{i}\langle{\phi}^{n-i-1}\rangle - 2\mg\varphi 
\langle XA\rangle - 2\mg \langle \phi XA\rangle = 0 && \nonumber \\
\hspace {12cm} \text{For (\eqref{decaymode})-b} && \label {dyneffb}
\eea
where $\phi$ is the quantum component of $\Phi$ with a null expectation value.
The next step in this calculation is determination of expectation values in 
\eqref{dyneffa} and \eqref{dyneffb}. The easiest method for this purpose is 
Schwinger closed time path (CTP) formalism. The detail of this calculation 
can be found in Ref.\cite{houridecohere}. Here we briefly discuss some of 
the results and prospects for future studies.

Considering only a mass term in the potential of $\Phi$ field, at radiation 
domination epoch after reheating the general solutions of the equations 
\eqref{dyneffa} and \eqref{dyneffb} at the WKB approximation level are:
\bea
\chi (\eta) &=& \chi_1 D_q (\alpha_\Phi \eta) + \chi_2 D_q (-\alpha_\Phi 
\eta) + 
g \int_{\eta_0}^\eta d\eta'  \langle \cx \ca^2\rangle G^\chi (\eta,\eta')
\quad \quad \text{For decay mode $(a)$} \label {gensola} \\
\chi (\eta) &=& \chi^{'}_1 D_q \biggl (\alpha_\Phi \int d\eta (1 - 
\frac {2g \langle \cx \ca\rangle}{a^2 m^2_\Phi})^{\frac{1}{4}}\biggr ) + 
\chi^{'}_2 D_q \biggl (-\alpha_\Phi \int d\eta (1 - 
\frac {2g \langle \cx \ca\rangle}{a^2 m^2_\Phi})^{\frac{1}{4}} \biggr ) + 
\nonumber \\
&& 2g \int_{\eta_0}^\eta d\eta' \langle \Upsilon \cx \ca\rangle 
G^\chi (\eta,\eta') \quad \quad \text{For decay mode $(b)$} \label {gensolb}
\eea
\bea
&& \alpha_\Phi \equiv (1+i)\sqrt{B_\Phi} \quad q \equiv - \frac{1+\frac{ik^2}
{B_\Phi}}{2} \quad B_\Phi \equiv \frac{a_0 m_\Phi}{\eta_0} = 
\frac {m_\Phi}{H_0 \eta_0^2} = a^2_0 H_0 m_\Phi \label{homosolparam} \\
&& \chi \equiv a\varphi \quad \Upsilon \equiv a\phi \quad \cx \equiv aX \quad 
\ca \equiv aA
\eea
where $\chi_1$, $\chi_2$, $\chi^{'}_1$ and $\chi^{'}_2$ and integration 
constants. $D_q$ is the parabolic cylindrical function. Expectation values in
\eqref{gensola} and \eqref {gensolb} include integration over multiple $D_q$ 
functions and don't have a closed analytical expressions. Using the asymptotic 
behaviour of $D_q (z)$ at large $z$, one finds that for both decay modes (a) 
and (b), the homogeneous part of the solutions \eqref{gensola} and 
\eqref{gensolb} is proportional to $\eta^{-1/2}$ and therefore deceases with 
time. 

In the same way, we can find the time dependence of the special solution. 
For large $\eta$ it is proportional to $\eta ^{2+\epsilon}$ for both decay 
modes. The parameter $\epsilon$ is added by hand to present the unknown time 
variation of the cosmological state $|\Psi\rangle$. Although in this formalism 
no explicit reference to time dependence of $|\Psi\rangle$ is made, its 
definition as the quantum state of particle in an expanding and evolving 
universe has implicitly the concept of time dependence in it. $\epsilon$ 
presents the back-reaction of interactions and expansion on the quantum 
state of the Universe. In classical treatment of particle production a 
dissipation term is usually added to express this 
back-reaction\cite{backreact}. In a field theoretical analysis of this 
problem the addition of a term by hand is not allowed and a more rigorous 
strategy must be taken.

In comoving coordinates the classical field $\varphi = \chi /a$ and varies as:
\bea
\varphi \propto t ^{-1/2+2+\epsilon} && \text{Assuming $m \neq 0$ for all 
fields, both decay modes.} \label{phiasympm} \\
\varphi \propto t ^{-1/2+3+\epsilon} && \text{For $m_{\Phi} \approx 0$ and 
mode (b).}\label{phiasympb}
\eea
For all cases, the late time behaviour of $\varphi$ depends on $\epsilon$
i.e. the back reaction and must be calculated numerically. We expect that 
$\epsilon$ depends on the density and lifetime of $X$ field, and on 
the expansion of the Universe. However, at late times when the energy 
density of the dark matter $X$ becomes comparable to the density of the dark 
energy $\sim 1/2 m_{\Phi}^2 \varphi^2$, there is a strong feedback between 
expansion rate and the density of dark matter - higher expansion brings down 
the density of the dark matter and therefore the rate of $\varphi$ production 
per unit of volume decreases. This reduces the density of the dark energy 
and the expansion. We have seen the same feedback in the classical 
formulation of this model in the previous section.

Finally it can be shown that $k$ dependence of $\varphi$ approaches a constant 
value at large scales consistent with observations of the dark energy.

\section{Conclusion}
The nature of dark matter and dark energy, along with search for a quantum 
theory of gravity are the greatest mysteries of contemporary physics. The can 
be related the first two can help - at least we hope - the third one which is 
more abstract and less available to cosmological observation or experiment in 
high energy accelerators.

Although numerous models for dark energy are proposed, the mere observation 
of its equation of state, whatever precise, can not distinguish between many 
of these models. Therefore one should find a relation between these models 
and other phenomena with independently quantities to be able to discriminate 
the right model.

Here we suggested a relation between dark matter and dark energy and fixed the 
properties of the model using completely independent observable: flux of the 
ultra high energy cosmic rays and supernovae observations. The closeness of 
predicted values in \eqref{wobs} to more recent observations is not a proof 
of the correctness of all aspects of this model. But maybe it is an 
indicator of the right way to proceed to the final solution.

\begin{theacknowledgments}
I would like to thank the organisers of Einstein 2005 Conference and more 
specifically J.M. Alimi. I would like also to thank D. Polarski and 
O. Bertolemi for constructive discussion during the conference. 
\end{theacknowledgments}

\bibliographystyle{aipproc}   

\bibliography{sample}

\IfFileExists{\jobname.bbl}{}
 {\typeout{}
  \typeout{******************************************}
  \typeout{** Please run "bibtex \jobname" to optain}
  \typeout{** the bibliography and then re-run LaTeX}
  \typeout{** twice to fix the references!}
  \typeout{******************************************}
  \typeout{}
 }


\end{document}